\documentclass[aps,prb,twocolumn,groupedaddress,superscriptaddress,showpacs]{revtex4}

\usepackage{amsmath}
\usepackage{graphicx}
\usepackage{bm}
\usepackage{hyperref}

\newcommand{\Mca}{\mathcal{M}}
\newcommand{\Hca}{\mathcal{H}}
\newcommand{\Kca}{\mathcal{K}}
\newcommand{\Pca}{\mathcal{P}}

\newcommand{\tp}{t_{\perp}}

\newcommand{\rA}{\text{A}}
\newcommand{\rB}{\text{B}}
\newcommand{\rD}{\text{D}}
\newcommand{\rND}{\text{ND}}

\newcommand{\e}{\epsilon}

\newcommand{\om}{\omega}

\newcommand{\vk}{{\bf k}}
\newcommand{\vx}{{\bf x}}
\newcommand{\vy}{{\bf y}}
\newcommand{\vq}{{\bf q}}
\newcommand{\vp}{{\bf p}}

\newcommand{\vpp}{{\bf p'}}
\newcommand{\vT}{{\bf T}}
\newcommand{\vK}{{\bf K}}

\newcommand{\vf}{v_{\rm F}}

\newcommand{\Ima}{{\rm Im}}
\newcommand{\Rea}{{\rm Re}}

\newcommand{\Hkin}{{\cal H}_{\rm kin}}
\newcommand{\Hex}{{\cal H}_{\rm I}}

\newcommand{\eo}{\epsilon_0}
\newcommand{\Eex}{E_{\rm ex}}
\newcommand{\Ekin}{E_{\rm kin}}
\newcommand{\Etot}{E_{\rm tot}}

\newcommand{\Qc}{Q_0}
\newcommand{\Qu}{Q_{\uparrow}}
\newcommand{\Qd}{Q_{\downarrow}}
\newcommand{\Qmin}{Q_{\rm min}}

\newcommand{\ordo}[1]{\mathcal{O}({#1})}
\newcommand{\chind}{\chi_{\rm ND} ( {\bf q} , \omega )}
\newcommand{\chid}{\chi_{\rm D} ( {\bf q} , \omega )}
\newcommand{\chirpa}{\chi_{\rm RPA} ( {\bf q} , \omega )}
\newcommand{\chifm}{\chi_{\rm FM} ( {\bf q} , \omega )}
\newcommand{\chiafm}{\chi_{\rm AFM} ( {\bf q} , \omega )}
\newcommand{\chio}{\chi_0 ( {\bf q} , \omega )}
\begin{document}

\title{Electron-electron interactions and the phase diagram of a graphene bilayer}

\author{Johan Nilsson}
\affiliation{Department of Physics, Boston University, 590 
Commonwealth Avenue, Boston, MA 02215, USA}

\author{A.~H. Castro Neto}
\affiliation{Department of Physics, Boston University, 590 
Commonwealth Avenue, Boston, MA 02215, USA}

\author{N.~M.~R. Peres}

\affiliation{Center of Physics and Departamento de F{\'\i}sica,
Universidade do Minho, P-4710-057, Braga, Portugal}

\author{F. Guinea}

\affiliation{Instituto de  Ciencia de Materiales de Madrid, CSIC,
 Cantoblanco E28049 Madrid, Spain}

\date{April 28, 2006}

\begin{abstract}
We study the effects of long and short-range electron-electron interactions in a graphene bilayer. Using a variational wavefunction technique we show that in the presence of long-range Coulomb interactions the clean bilayer is always unstable to electron {\it and} hole pocket formation with a finite ferromagnetic polarization. Furthermore, we argue that short-range electron-electron interactions lead to a staggered orientation of the ordered ferromagnetic moment in each layer (that is, c-axis antiferromagnetism). We also comment on the effects of doping and trigonal distortions of the electronic bands.
\end{abstract}


\maketitle

\section{Introduction}

The recent developments in the field of carbon physics, where
a few layers or even single layers of graphene have been isolated, have shown that the physics of these systems is unconventional from the point of view of traditional semiconductor and Fermi-liquid
physics \cite{Novolelov2005,Zhang2005}. The electronic dispersion of
graphene close to the two K-points of the Brillouin zone can be written as \cite{Wallace47}:  $E_{\pm}({\bf p}) = \pm v_F |{\bf p}|$,
where $v_F$ is Dirac-Fermi velocity (this expression is valid for two-dimensional momentum ${\bf p}=(p_x,p_y)$ such that $|{\bf p}|<\Lambda$ where $\Lambda$ is a momentum cut-off of the order of the inverse of the lattice spacing $a$). This dispersion relation
is identical to the one of Dirac electrons with ``speed of light"
given by $v_F$.  In this case the electron effective mass, $m^*$, is zero, and the density of states vanishes at the K-point. The vanishing of the effective mass, the interplay of interactions, disorder, and extended defects, lead to anomalous behavior in many physical properties \cite{nuno_curto,nuno2006_long}. 

The capability of experimentally controlling the number of graphene
layers opens up the field for the study of the effect of interlayer coupling in a strongly interacting two-dimensional system. Interlayer coupling is a controversial topic in the 
graphite literature where the precise nature of the coupling between graphene planes is unsettled \cite{BCP88,Rydberg2003}. Another important issue in carbon research has to do with the weak ferromagnetism in highly disordered graphite that have been observed in experiments \cite{esquinazi} but is still a theoretically open problem \cite{Vetal05,Peres2005_ferro}.

It is well-known that the low-density electron gas with long-range Coulomb interactions in two and three
dimensions is unstable toward a ferromagnetic state.
The original argument due to Bloch 
relies on a variational calculation of the ground state 
energy \cite{Bloch}. Recently this approach 
was used to look for a possible ferromagnetic instability in a
single layer of graphene \cite{Peres2005_ferro}. The parameter
that controls the relative strength between kinetic and Coulomb
energies is the dimensionless coupling, 
$
g = e^2 \epsilon^{-1}_0/v_F \, ,
$
($\hbar =1$) where $e$ is the electric charge ($e^2 = 14.4$ eV \AA), and 
$\epsilon_0$ is the graphene dielectric constant ($\epsilon_0 \approx 1$). 
In that case, ferromagnetism is only found for values of $g$ larger
than a critical value, $g_c \approx 5.3$, which is larger than
its estimated value in graphene ($g \approx 2.1$). 
An analysis based on short-range interactions seems to confirm this picture \cite{nuno2006_long}. 

In this paper we use a similar variational technique to study a clean graphene bilayer 
where we include the hopping between graphene planes.
Unlike the case of a single layer, we find that the bilayer is always unstable toward a ferromagnetic state with formation of electron and hole pockets with a polarization of the order of $10^{-6}$ to $10^{-5}$ electrons per carbon. This result may have direct implications for the interpretation of the magneto-transport data in graphitic devices \cite{berger04}.

The paper is organized as follows: In Section \ref{sec_Model} the
model is introduced. In Section \ref{sec:variational} 
we explain the variational
calculation and present the phase diagram.
The influence of other hopping parameters on the instability
are discussed in Section \ref{sec:other_hopping}.
Section \ref{sec:shortrange} includes the results for the
low-energy susceptibilities and a
discussion of short range interactions.
The conclusions of the paper are to be found in Section \ref{sec:conclusions}. 
We also include appendices with some mathematical details.

\section{The model}
\label{sec_Model}
\begin{figure}[htb]
\includegraphics[scale=0.5]{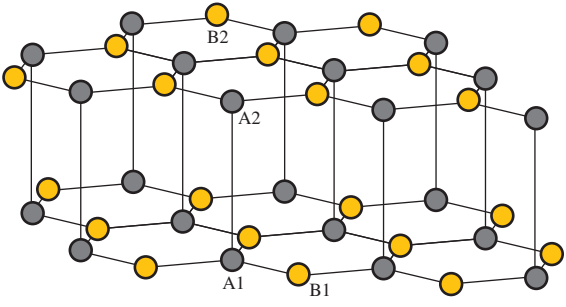}
\caption{(Color online) Lattice structure of the bilayer. The
  A-sublattices are indicated by the darker spheres.}
\label{fig_lattice}
\end{figure}
The lattice structure for the bilayer which is just one unit cell of graphite
is depicted in Fig. \ref{fig_lattice}.
For simplicity we model the system by the nearest neighbor
tight-binding Hamiltonian:
\begin{eqnarray}
  \label{eq:HTB}
  \Hca_{\text{t.b.}} &=& 
  -t \sum_{\substack{<m,n> \\ i, \sigma}}
  (c_{A_i,m,\sigma}^{\dag} c_{B_i,n,\sigma}^{\,} + \text{h.c.} )
  \nonumber \\
  &-& \tp \sum_{m,\sigma} (c_{\rA_1,m,\sigma}^{\dag} 
  c_{\rA_2,m,\sigma}^{\,} + \text{h.c.}),
\end{eqnarray}
where $c_{a_i,m,\sigma}^{\,} \,(c_{a_i,m,\sigma}^{\dag})$ annihilates
(creates) an electron on site $m$ of the sublattice $a$ ($a=A,B$) of
plane $i$ ($i=1,2$), with spin $\sigma$ ($\sigma = \uparrow,\downarrow$),
$t$ ($t \approx 3$ eV) is the in-plane hopping energy and 
$\tp$ ($\tp \approx 0.35$ eV in graphite\cite{BCP88}) 
is the hopping energy between atom $\text{A}_1$ and atom $\text{A}_2$ 
(see Fig.~\ref{fig_lattice}).
A similar tight-binding Hamiltonian for graphite and the single graphene layer
was studied long ago by Wallace\cite{Wallace47}.
At low energies and long wavelengths, the kinetic Hamiltonian can
be expanded around the K (K') points in the Brillouin zone.
The resulting Hamiltonian can be written as: 
\begin{equation}
  \label{eq:10}
  \Hkin = \sum_{{\cal Q}} \Psi^{\dag}_{{\cal Q}} \Kca( \vp ) \Psi_{{\cal Q}},
\end{equation}
with ${\cal Q}$ denoting $(\vp , \alpha_i,\sigma, a) $ and 
$\Psi^{\dag}_{{\cal Q}}  =
( c_{\vp, rA_1, \sigma, a}^{\dag} , \:
  c_{\vp, rB_1, \sigma, a}^{\dag} , \:  
  c_{\vp, rA_2, \sigma, a}^{\dag} , \: 
  c_{\vp, rB_2, \sigma, a}^{\dag})$. 
Here, $c^{\dag}_{\vp,\alpha_i,\sigma,a}$ ($c_{\vp,\alpha,\sigma,a}$) creates
(annihilates) an electron with momentum $\vp$, on sublattice
$\alpha_i$ ($\alpha=\rA,\rB$) of plane $i$ ($i=1,2$), with spin $\sigma$ 
($\sigma=\uparrow,\downarrow$) at the K-point $a$ ($a=1,2$) in
the Brillouin zone, and
\begin{equation}
  \label{eq:Hkin0}
 \Kca(\vp) = 
  \begin{pmatrix}
    0 & p e^{i \phi(\vp)} & -\tp & 0 \\
    p e^{-i \phi(\vp)} & 0 & 0 & 0 \\
    -\tp & 0 & 0 & p e^{-i \phi(\vp)} \\
    0 & 0 & p e^{i \phi(\vp)} & 0
  \end{pmatrix} \, ,
\end{equation}
is the kinetic energy matrix where $\phi(\vp) = \tan^{-1}(p_y/p_x)$. 
We have set $v_F = 1 = \Lambda$, so that the energy is measured in
units of the in-plane hopping, $t$ , 
and distance is measured in units of carbon-carbon distance $a$ ($a \approx 1.42$ \AA) \cite{BCP88}. 
 
\begin{figure}[htb]
\includegraphics[scale=0.5]{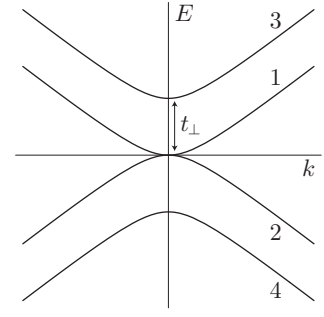}
\caption{Band dispersions near the K-points in the
  bilayer. Bands are labeled by the numbers $1-4$ as in the text.}
\label{fig_bands}
\end{figure}

The kinetic term can be diagonalized by a unitary transformation: 
$\Psi_{\vp} = \Mca_{\vp} \Phi_{\vp}$, where
$\Mca_{\vp}$ is given in Appendix~\ref{app:unitary}.
Then
$\Hkin = \sum_{\vp,j,\sigma,a} E_{j} (p)\Phi^{\dag}_{\vp,j,\sigma,a}
 \Phi_{\vp,j,\sigma,a}$, where the four energy
bands are given by:
\begin{eqnarray} 
E_{1}(p) &=& -\tp/2 + E(p) \, , 
\nonumber
\\
E_{2}(p) &=& \tp/2 - E(p) \, ,
\nonumber
\\
E_{3}(p) &=& \tp/2 + E(p) \, ,
\nonumber
\\
E_{4}(p) &=& -\tp/2 - E(p) \, ,  
\nonumber
\end{eqnarray}
where $E(p) = \sqrt{\tp^2/4 + p^2}$. The bands are
sketched in Fig.~\ref{fig_bands}.
Any state of the system can be
labeled in terms of the occupation of each band, 
$n_{i,\sigma,a}(p)$, with $i=1,2,3,4$.
The non-interacting ground state has degeneracy $4$ per momentum, 
 per plane, due to the SU(2) spin rotation symmetry and the Z$_2$ 
real space sublattice exchange symmetry (at low energies this symmetry becomes
SU(2) for the continuous rotation of the K and K' states in momentum space), 
and occupation at half-filling, given by: 
$n_{1,\sigma,a}(p) = 0$, $n_{2,\sigma,a}(p) = 1$, 
$n_{3,\sigma,a}(p) = 0$ and $n_{4,\sigma,a}(p) = 1$. Hence,
the presence of $\tp$ does not mix the spins or the K-points. However, the two
Dirac cones transform into vertex touching hyperbolae, 
and for $p \ll \tp$ the electrons acquire an effective mass, 
$m^* \approx \tp/2$. 

The Coulomb interaction in the bilayer is conveniently written
in terms of the Fourier components of
symmetric and anti-symmetric combinations of the layer densities,
$\rho_{\pm}(\vq) = \rho_{1}(\vq) \pm
\rho_{2}(\vq)$, where $\rho_i(\vq) = \sum_{\vk,\alpha,\sigma,a}
c^{\dag}_{\vk+\vq,\alpha_i,\sigma,a} 
c_{\vk,\alpha_i,\sigma,a}$. The Coulomb term reads: 
\begin{equation}
  \label{eq:2planecoulomb_pm}
  \Hex = \frac{1}{2 S} \sum_{\bm q \neq 0} \sum_{\alpha=\pm}
  \rho_{\alpha}(\vq) V_{\alpha}(\vq) \rho_{\alpha}(-\vq),
\end{equation}
where $V_{\pm}(\bm{q}) = 2 \pi e^2(1 \pm e^{-q d})/ 2 \eo q$, $S$
is the area of the system, and $d$ is the interplane distance ($d \approx 2.4 a \approx 3.35 \text{\AA}$, $\Lambda d \approx 3.7$).
We are going to show that in the presence of Eq.~(\ref{eq:2planecoulomb_pm}) 
the non-interacting ground state is unstable. 
To perform the calculation it is convenient to express the density
operators in the diagonal basis:
$\rho_{\pm}(\vq) = \sum_{\vp,i,j,\sigma,a} \Phi_{\vp + 
\vq,i,\sigma,a}^{\dag} 
   \chi_{ij}^{\pm}(\vp + \vq,\vp) \Phi_{\vp,j,\sigma,a}$
   and write the exchange energy
   associated with Eq.~(\ref{eq:2planecoulomb_pm}) as:
\begin{eqnarray}
  \label{eq:Hexvar}
  \frac{\Eex}{S} = - \frac{1}{2} 
  \int \frac{d \vp}{(2 \pi)^2} \frac{d^2 \vpp}{(2 \pi)^2}
  \sum_{\alpha=\pm ,i,j,\sigma,a} \nonumber \\
  \chi^{\alpha}_{ij}({\bf p'},{\bf p})
  \chi^{\alpha}_{ji}({\bf p}, {\bf p'})
  n_{i,\sigma,a}(p') n_{j,\sigma,a}(p) V_{\alpha}({\bf p'} - {\bf p}).
\end{eqnarray}
The definitions of the matrices $\chi^{\pm}$ and some more details
about the exchange interaction for Bloch electrons and 
Eq.~(\ref{eq:Hexvar}) are given in Appendix~\ref{app:Blochexchange}.

\section{Variational calculation and Phase diagram}
\label{sec:variational}
\begin{figure}[htb]
\includegraphics[scale=0.22]{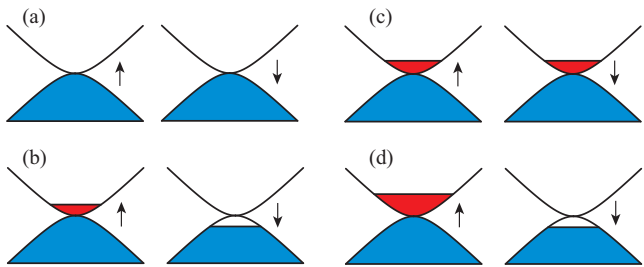}
\caption{(Color online) Sketch of the trial states: 
  {\bf (a)} Half-filled non-interacting ground state,
  {\bf (b)} Trial state with particle-hole pockets built upon {\bf (a)},
  {\bf (c)} Doped non-interacting ground state,
  {\bf (d)} Trial state with particle-hole pockets built upon {\bf (c)}.
}
\label{fig_trial}
\end{figure}

Consider the half-filled case with a variational state with one electron pocket in the spin up channel and one hole pocket in 
the spin down channel at each K-point: 
$n_{1,\uparrow,a}(p) = \Theta(Q - p)$, $n_{1,\downarrow,a}(p) = 0$, $n_{2,\uparrow,a}(p) = 1$,  and $n_{2,\downarrow,a}(p) =1-\Theta(Q - p)$,
where $Q$, the size of the pocket, is a variational parameter (in what follows we assume $Q \ll \tp$ and hence the occupations of bands $3$
and $4$ are not affected). Pictures of the non-interacting ground
state and the trial state are shown in Fig.~\ref{fig_trial}a 
and \ref{fig_trial}b.
Notice that the size of the pocket is the same in
different channels because of the conservation of the number of electrons at
half-filling. This state breaks the SU($2$),
but not the Z$_2$ symmetry, and is therefore spin polarized (ferromagnetic). 
There is a similar state that breaks both symmetries and has no
net magnetization: an electron (hole) pocket in the up (down) spin channel in K-point $1$
and a hole (electron) pocket in the up (down) channel in K-point $2$. We can show
that the spin polarized state is lower in energy (see below).

The change in the kinetic energy per unit area due to an electron
 (or hole) pocket of size
$Q$ is given by:
\begin{equation}
  \label{eq:DeltaEkin}
  \frac{\Delta \Ekin}{S} = 
  \frac{1}{2 \pi}\bigr[\frac{(Q^2 + \tp^2 / 4)^{3/2}}{3} 
  - \frac{\tp^3}{24}
  - \frac{\tp Q^2}{4} \bigr]
  \approx \frac{Q^4}{8 \pi \tp} ,
\end{equation}
up to order $Q^4$.
The expressions for the change in the exchange energy are cumbersome and 
details are provided in Appendix \ref{app:exchange_integral}. 
We find from Eq.~(\ref{eq:Hexvar}), up to the same order:
\begin{multline}
  \label{eq:effectiveenergy1}
  \frac{\Delta \Eex(Q)}{S}
\approx
- \frac{g}{8 \pi^2} \biggr\{
\frac{8}{27} \, Q^3 -
\Bigl[  -  \frac{3 \pi}{8} 
\\ 
 + 
\frac{\pi}{2}  \ln(4 / \tp)
  -\frac{\pi}{4}\int_0^1 dy \frac{e^{-d \Lambda y}}{\sqrt{\tp^2 /4 +y^2}}
\Bigr] \frac{Q^4}{\tp} \biggr\}. 
\end{multline}
Notice that the leading order term in the exchange interaction
is $ \Delta \Eex /S \sim -  g Q^3 / 27 \pi^2 \sim - m^{3/2}$, 
where $m$ is the magnetization, which is always
dominant over the kinetic term that is of order $Q^4 \sim m^2$.
Therefore, we have proved that the bilayer is {\it always
unstable} to the formation of polarized electron and hole
pockets. In contrast to the single graphene plane case \cite{Peres2005_ferro},  
the total energy is negative for small $Q$. This is due to the
fact that the exchange with the filled bands is less important in this
case. In order to calculate the equilibrium size of the pockets we minimize
the total energy, $\Delta E_{{\rm tot}}(Q) = \Delta \Ekin(Q) + \Eex(Q)$,
with respect to $Q$ and find $\Qmin$, that is, the size of the pocket for which the energy is minimized. For the parameters in graphene (see below)
we find that $\Qmin \sim 0.05 \tp$ ($\ll \tp$), justifying the above expansion. 

\begin{figure}[htb]
\includegraphics[scale=1]{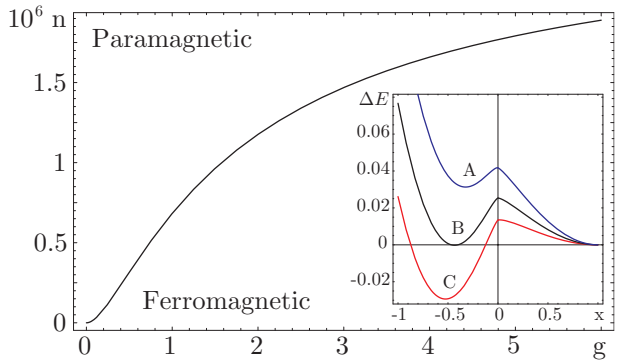}
\caption{(Color online) Left: Phase diagram of the graphene bilayer as a
  function of electron density away from half-filling, $n$
(electrons per carbon), and coupling strength, 
$g=e^2/(\epsilon_0 v_F)$, with $\tp = 0.05$.
Inset: $\Delta E$ as a function of $x$ (as defined in the text) in the paramagnetic (A), critical (B), and ferromagnetic (C) regions of the phase diagram.}
\label{fig_phasediagram}
\end{figure}

Consider the case where  the system is initially doped with pockets of size $\Qc$  ($n=\Qc^2/2$). We look for an instability by varying the density of electrons and holes
subject to the constraint of particle conservation. Note that the instability can
produce one type of carrier (either electron or hole) if $\Qc > \Qmin$ or two
type of carriers (electrons and holes) if $\Qc < \Qmin$. 
We can parameterize the state with one type of carrier 
by taking $\Qu^2 = \Qc^2 \, (2 - x)$ and $\Qd^2
= \Qc^2 \, x$ with $0 \leq x \leq 1$.
For the state with two types of carriers we take instead 
$\Qu^2 = 2 \Qc^2 +|x|$ and $\Qd^2 = |x|$, with $x \leq 0$.
The doped non-interacting ground state and the trial state with particle-hole
pockets ($x<0$) are shown pictorially in Fig.~\ref{fig_trial}c and
\ref{fig_trial}d.
The calculation proceeds as before and we find:
\begin{eqnarray}
  \label{eq:DeltaEdoped}
  \frac{\Delta E}{S} \approx
  \frac{1}{2 S}\bigl[\Delta \Etot(\Qu)+\Delta \Etot(\Qd)
  -2 \Delta \Etot(\Qc)\bigr]
\nonumber \\
  + \Delta E_{{\rm extra}}(\Qc,x).
\end{eqnarray}
The extra term, $\Delta E_{{\rm extra}}(\Qc,x)$, comes from terms that cancel out in the
undoped case. To leading order in $Q$ these terms are given 
by $g \, \Qc^4 (1-x)^2 /
(64 \pi)$ for $x \geq 0$ and $g \, \Qc^2 (\Qc^2 + 2 |x|) /
(64 \pi)$ when $x \leq 0$.
In our units we have $\tp  \ll 1$ so that, to a first approximation, this
contribution is much smaller than the quartic term in 
Eq.~(\ref{eq:effectiveenergy1}), and it can be neglected.
This leaves us with the first line in Eq.~(\ref{eq:DeltaEdoped}) involving
$\Delta E_{\rm tot}$ only. The dependence on $x$ is implicit through
$\Qu$ and $\Qd$. 
Then Eq.~(\ref{eq:DeltaEdoped}) has the form: 
$\Delta E(Q) = -A |Q|^3 + B |Q|^4$. Rescaling the
$Q$ variable so that the minima of the energy in the
paramagnetic states sits at $|Q|=1$, we have: 
\begin{eqnarray}
\Delta E(Q) = - |Q|^3 /3 + |Q|^4 /4.
\end{eqnarray}
Using the scaled variables we see that the system is unstable
to small deviations in $x$ from $1$ if  $\Qc \leq 1/2$. The ferromagnetic state has lower energy than the paramagnetic
states if $Q_0 \lesssim 0.7$ and the resulting state 
has electron and hole pockets. As a consequence of the first
order nature of the transition, the system exhibits phase coexistence
(that can be obtained from a Maxwell construction, not
shown in the Fig.~\ref{fig_phasediagram}), 
and hysteresis in physical properties such as
magneto-transport, around the critical line.  In this region, the system shows a tendency towards electronic phase separation \cite{GGA00}, frustrated by electrostatic effects. As
the charge densities involved are rather low (see below) we cannot exclude the formation of large domains of the different phases.
The phase diagram for $\tp = 0.05$ 
is shown in Fig. \ref{fig_phasediagram} as well as
a plot of $\Delta E(\Qc,x)$ for some typical cases of $\Qc$.

In the previous calculation we have not included the exchange interaction
between different K-points in the Brillouin zone. In that case the
spin polarized state that breaks SU($2$) is degenerate with the state
that breaks both SU($2$) and Z$_2$. The difference between the states is how 
the pockets are assigned to the spins and the K-points. 
By including exchange between K-points in Eq.~(\ref{eq:Hexvar})
we find that there is small energy difference between the states favoring a 
state with a net ferromagnetism but which retains the Z$_2$ symmetry. 
Quite generally this is the case since the elements of
$\chi^{\alpha}_{ij}({\bf p'},{\bf p}) \chi^{\alpha}_{ji}({\bf p}, {\bf
  p'})$ are all positive.
A direct calculation using Eq.~(\ref{eq:Hexvar}) and taking $\vp$ and
$\vpp$ to lie at nearest neighboring K-points confirms this picture.
One then finds a very small energy difference of order $\propto Q^4$,
and hence other effects can be important in  determining the actual
ground state.
There is also a correction to the
$Q^4 / \tp$-term in Eq.~(\ref{eq:effectiveenergy1}) 
that changes the position of the optimal value of $\Qmin$ by a small amount.

In order to compare with experiments it is interesting to estimate
the total magnetization in the polarized state for the case of the undoped
graphene bilayer. We estimate the cut-off using a Debye approximation
so the the number of states is conserved in the Brillouin zone: 
$\Lambda^2 = 2 \pi/A_u = 4 \pi/(\sqrt{27} a^2)$, where $A_u$ is the area of the real space unit cell. Restoring the units, 
we set $v_F = 3 \gamma_0 a/2 \approx 10^6$ m/s, and $\tp=\gamma_1/( v_F \Lambda) \approx 5.2 \times 10^{-2}$, where $\gamma_1 \approx 0.37$ eV is the typical graphite value \cite{BCP88}.
Hence, for two pockets of size $Q$ the density of electrons per carbon is
approximately  $n=Q^2 / 4 \approx 1.6 \, \times \, 10^{-6}$ 
($\tp \sim 0.05$ and $Q \sim 0.05 \tp$), 
and therefore, 
the magnetization per carbon is $m \approx 10^{-6} - 10^{-5} \mu_B$ 
($\mu_B$ is the Bohr magneton). These number are, of course, very approximate
because the value of the microscopic parameters do not need to be the same as
in graphite and the presence of a cut-off introduces further uncertainty. In
any event, the magnetized state of the graphene bilayer shows very weak ferromagnetism.
A direct experimental consequence of our calculation is that the bilayer has two species
of electrons (electrons and holes) and therefore they should contribute to the Hall 
resistivity at small magnetic fields, $B$. In particular, it is easy to show that the
magneto-resistance at small magnetic fields acquires a $B^2$ dependence \cite{Duetal04}. 

\section{Other hopping parameters}
\label{sec:other_hopping}
We would like to comment on other effects that we have
not considered in the previous calculation. In terms of the Slonczewski-Weiss-McClure
model for graphite \cite{M57,SW58} our
model includes only the parameters $\gamma_0$ ($t$) and $\gamma_1$ ($\tp$)
but not $\gamma_3$ and $\gamma_4$. On the one hand, $\gamma_4$ introduces
an electron-hole asymmetry by changing the curvature of the bands, but
the bands remain parabolic near half filling. On the other hand, 
$\gamma_3$ introduces a trigonal distortion which restores a linear dispersion at low energies.
To estimate the effects of $\gamma_3$ we use the effective low-energy
model that can be derived from the extension of Eq.~(\ref{eq:Hkin0})
to include $\gamma_3$ and
projecting onto the two bands that are closest to the Fermi
surface (see Ref.~[\onlinecite{Falko2006}] and Appendix \ref{app:effective}
for details). The effective kinetic energy matrix is then:
\begin{equation}
  \label{eq:Ewarping}
  {\cal K}({\bf p}) = \frac{p^2}{\tp}
  \begin{pmatrix}
    0 & e^{- i 2 \phi} \\
    e^{i 2 \phi} & 0
  \end{pmatrix}
  +
  v_3 p
  \begin{pmatrix}
    0 & e^{i \phi} \\
    e^{-i \phi} & 0
  \end{pmatrix},
\end{equation}
where $v_3 = \gamma_3 / \gamma_0$, with energy eigenvalues given by:
\begin{eqnarray}
E_{\pm}(\vp) = \pm p \sqrt{p^2+ (v_3 \tp)^2+ 2 p  v_3 \tp  \cos[3\phi(\vp)]
}/ \tp.
\end{eqnarray}
It is easy to see that the
crossover from linear to quadratic dispersion takes place at a momentum 
$p_{{\rm cross}} \sim \, v_3 \, \tp$. If $p_{{\rm cross}} \ll Q_{{\rm min}}$ ($v_3 \ll 0.1$) 
the previous calculation is still valid since the dispersion remains parabolic
at the scale of the instability. Nevertheless, if one uses the values of the
parameters in the graphite literature \cite{BCP88}, namely,  
$v_3 = \gamma_3 / \gamma_0 \approx 0.1$
we conclude that $p_{{\rm cross}} \sim Q_{{\rm min}}$ and the trigonal
distortion may become important. We should remark, once again, that it is
not guaranteed that the value of the parameters in graphite are the same as in
the bilayer (hence, one should take the numbers here with a grain of salt).
At small energies, the spectrum in Eq.~(\ref{eq:Ewarping}) can be described in terms of 
one Dirac cone at $p=0$, and three asymmetric cones at $p_{{\rm cone}}= v_3
\tp$ (the direction of the cones are such that $\cos(3 \phi) = -1$). This situation
maps onto the single graphene plane case \cite{Peres2005_ferro} with ellipsoidal
pockets instead of circular, and a small renormalized Fermi velocity, 
$v_F^* \approx v_3 v_F$ ($\ll v_F$). Therefore the dimensionless
coupling strength is
$g^*=g/v_3 \approx 20$, which is much larger than the critical coupling for ferromagnetism \cite{Peres2005_ferro}. Then, in this case the transition from
paramagnetism is into a fully polarized state with effective bandwidth
of order of $v_3 \tp$, leading to a polarization of the order of
$10^{-6}$ electrons per carbon, which is of the same order of
magnitude of polarization found without $\gamma_3$.
Unfortunately this argument is not rigorous since the exchange with
the filled bands also contributes, requiring a more detailed study.

\section{Short range interactions}
\label{sec:shortrange}
We consider the effects of short-range
electron-electron interactions which, for simplicity, we describe by
an  on-site Hubbard interaction, $U$.
It turns out that this interaction favors an antiferromagnetic
ordering.
In order to quantify the tendency towards this phase, we
calculate the associated susceptibility and present a simple mean
field argument.

\subsection{Electronic susceptibility.}
Using the basis defined in Eq.~(\ref{eq:Ewarping}) with $v_3 =0$ 
(the procedure outlined in Appendix \ref{app:effective})
 the wavefunctions corresponding
to the two bands closest to the Fermi level can be written approximately as:
\begin{equation}
\Psi_{\pm} ( \vk ) \equiv \frac{1}{\sqrt{2}}\left( \begin{array}{c}
    e^{-i \phi(\vk)} \\
    \pm e^{i \phi(\vk) } \end{array} \right) \, ,
\end{equation}
    so that the states are mostly localized at the sites without a
    corresponding atom in the neighboring layer. Formally, this two component
    wavefunction is equivalent to the spinor defined in the analysis of a
    single graphene plane except that the angle is doubled.
 Restricting the calculation to this subspace, we
    can write the bare susceptibility as a $2 \times 2$ matrix
(Ref.~\onlinecite{Peres03}):
\begin{equation} \chio = \left( \begin{array}{cc}
\chid & \chind \\ \chind & \chid \end{array} \right).
\label{susc_0} \end{equation}
Here $\chi_{\rm D}$ denotes the response in the same plane as the
source and $\chi_{\rm ND}$ response in the opposite plane.
The random phase approximation (RPA) susceptibility, assuming an on-site Hubbard interaction $U$, is:
\begin{equation}
  \chirpa = \chio \bigl[ 1 - U \chio \bigr]^{-1} \, .
  \label{susc_RPA}
\end{equation}
This expression becomes simpler
when decomposed into a contribution symmetric in the two
sublattices and another antisymmetric: 
\begin{eqnarray} 
\frac{\chifm}{4} &= &\frac{\chid + \chind}{1 - U \left[ \chid + \chind
  \right]} \, ,
\nonumber \\ 
\frac{\chiafm}{4} &= &\frac{\chid - \chind}{1 - U \left[ \chid
- \chind \right]} \, . 
\end{eqnarray}
The symmetric susceptibility
gives the response of the system to a magnetic field which is the
same in the two sublattices (note that we are neglecting the influence of the
sites where the states have zero weight of $\vk = 0$), and induces a ferromagnetic ordering.
The antisymmetric response leads to antiferromagnetic ordering.
The susceptibilities can be written as:
\begin{widetext}
\begin{eqnarray}
 \chid &= & - \frac{\tp}{4 \pi}
\int_0^\Lambda k d k \int_0^{2 \pi} \frac{d \phi}{2 \pi} \, \,
\left[ \frac{1}{\omega \tp -  | \vk + \vq / 2 |^2 - | \vk - \vq / 2 |^2 }
- \frac{1}{\omega \tp +  | \vk + \vq / 2 |^2 + | \vk - \vq / 2 |^2 } \right]
 \nonumber \\ 
 \chind &= &  \frac{\tp}{4 \pi}
\int_0^\Lambda k d k \int_0^{2 \pi} \frac{d \phi}{2 \pi} 
\left[
\frac{\cos \left( 2 \phi_{ \vk + \vq / 2, \vk - \vq / 2} \right)}
{\omega \tp - | \vk + \vq / 2 |^2 - | \vk - \vq / 2 |^2 } 
-
\frac{\cos \left( 2 \phi_{ \vk + \vq / 2, \vk - \vq / 2} \right)}
{\omega \tp + | \vk + \vq / 2 |^2 + | \vk - \vq / 2 |^2 } 
\right],
\label{eq:susc_integrals1}\end{eqnarray} 
where $\phi_{ \vk + \vq / 2, \vk - \vq / 2}$ is the angle between $\vk
+ \vq / 2$ and $\vk - \vq / 2$,
and:
\begin{equation}
  \cos^2 \left( \phi_{ \vk + \vq / 2, \vk - \vq / 2} \right)
  = \left[ \frac{( \vk + \vq / 2 ) \cdot ( \vk - \vq / 2 )}
    {|\vk + \vq / 2 | | \vk - \vq / 2 |} \right]^2 
  = \frac{( k^2 - q^2 / 4 )^2}
  {(k^2 +q^2 /4)^2 - k^2 q^2 \cos^2 ( \theta )}
\label{cosine}
\end{equation}  
The only dependence on the angle $\theta$ between the vectors $\vk$ and $\vq$
of the expressions in Eq.~(\ref{eq:susc_integrals1}) is (after using
the double angle formula) through the cosine in
Eq.~(\ref{cosine}). Averaging over angles, we obtain:
\begin{equation}
\left\langle \cos^2 \left( \phi_{ \vk + \vq / 2, \vk
  - \vq / 2} \right) \right\rangle = 
\left|\frac{ | \vk |^2 - | \vq |^2 / 4}{ | \vk |^2 + | \vq |^2 / 4} \right|
\, .
\end{equation}
Inserting this expression into Eq.~(\ref{eq:susc_integrals1}) it is a
simple task to perform the remaining one-dimensional integral. 
Introducing $\chi_{\rm FM}^0 = \chid + \chind$ and 
$\chi_{\rm AFM}^0 = \chid - \chind$
we can extract the leading dependence on the cut-off
$\Lambda$ of the susceptibilities, and we finally obtain:
\begin{eqnarray}
  \Rea \chi_{\rm FM}^0 ( \vq , \omega ) &= &
 \frac{1}{4 \pi} \sum_{s=\pm 1}
    \Bigl[\tp \ln \Bigl| 
       \frac{| \vq |^2 - s \om \tp}{| \vq |^2 / 2 - s \om \tp}
    \Bigr|
  +\frac{| \vq |^2}{2 s \om} 
  \Bigr(
  \ln \Bigl| \frac{2 | \vq |^2}{| \vq |^2/2 - s \om \tp} \Bigr|
  -2 \ln \Bigl| \frac{| \vq |^2 - s \om \tp}{| \vq |^2/2 - s \om \tp}
  \Bigr|
  \Bigr)
  \Bigr] \, ,
\nonumber \\
\Rea \chi_{\rm AFM}^0 ( \vq , \omega ) &= &
\frac{\tp}{8 \pi} \sum_{s =\pm 1} 
\ln \Bigl| \frac{2 \Lambda^2}{|\vq |^2/2 - s \om \tp} \Bigr|
- \Rea \chi_{\rm FM}^0 ( \vq , \omega ) \, ,
\nonumber \\
\Ima \chi_{\rm FM}^0 ( \vq , \omega ) &= &
\frac{\tp}{4} \Bigl[
 \Bigl(1 -\frac{|\vq|^2  }{2 |\om| \tp} \Bigr) 
 \theta \Bigl( |\omega| - \frac{| \vq |^2}{2 \tp} \Bigr)
- \Bigl(1 -\frac{|\vq|^2  }{|\om| \tp} \Bigr) 
 \theta \Bigl( |\omega| - \frac{| \vq |^2}{\tp} \Bigr)
 \Bigr] \text{sign}(\om) \, ,
  \nonumber \\ 
\Ima \chi_{\rm AFM}^0 ( \vq , \omega ) &= & 
 \frac{\tp}{8}
\theta \Bigl( |\omega| - \frac{| \vq |^2}{2\tp} \Bigr)\text{sign}(\om)
-\Ima \chi_{\rm FM}^0 ( \vq , \omega ) \, .
\label{chi} \end{eqnarray}
\end{widetext}
Hence, setting $\vq = 0$, the antiferromagnetic susceptibility diverges
logarithmically with the cut-off for any finite frequency $\omega$. 
The logarithmic dependence implies the existence of an
instability for any positive value of the interaction $U$. Alternatively, we
can show the existence of this instability by a direct calculation of the
correlation energy gained by polarizing the system.

It is
worth noting that the divergence obtained here, and the related marginal
behavior of a local interaction can be obtained from the same power counting
arguments used for the analysis of two dimensional interacting electrons near
a van Hove singularity\cite{GGV97b}.
If we include next nearest neighbor couplings through the parameter
$\gamma_3$, as discussed in Section~\ref{sec:other_hopping}, 
the low energy bands can be described by an effective Dirac
equation. The screening of the long range Coulomb interaction
vanishes, 
and the corresponding susceptibility can also be calculated 
analytically\cite{GGV94b,GGV96}. 

Note also that the polarization function is simply related to the
susceptibilities\cite{mahan} above by $\Pi^0 = - \chi^0$.
This allows one to get
the screening properties within the RPA easily. 
In particular, for the mode that
is symmetric in the layer densities which originally had the
long-wavelength $1/|\vq|$ singularity the static ($\om =0$) RPA 
screening cuts off the singularity by
taking $1/|\vq| \rightarrow 1/(|\vq| + q_{\text{TF}})$.
From Eq.~(\ref{chi}) we find that the Thomas-Fermi screening 
wavevector $q_{\text{TF}} \propto \tp$.
This is in agreement with
what one expects from the usual two-dimensional electron
gas where the screening wave-vector 
is $\propto m^*$ independently of the
density of carriers\cite{Canel1972,Fetter1973}.
\subsection{Mean-field approach}
Alternatively we can explain the diverging susceptibility  with a
simple mean-field approach. We introduce a staggered
mean-field $\Delta$ into the Hamiltonian according to
\begin{equation}
  \label{eq:Hkin0MF}
 {\cal K}({\bf p}) = 
  \begin{pmatrix}
    \pm \Delta & p e^{i \phi(\vp)} & \tp & 0 \\
    p e^{-i \phi(\vp)} & \mp \Delta & 0 & 0 \\
    \tp & 0 & \mp \Delta & p e^{-i \phi(\vp)} \\
    0 & 0 & p e^{i \phi(\vp)} & \pm \Delta
  \end{pmatrix} \, ,
\end{equation}
where the $\pm$ label different spin orientations.
The four bands are then:
\begin{equation}
  \label{eq:MFbands}
  \e(\vk) = 
  \pm \sqrt{\frac{2 k^2 + \tp^2  + 2\Delta^2 \pm \tp \sqrt{4\ k^2 + \tp^2} }{2}}.
\end{equation}
The opening of the gap lowers the kinetic energy of the system.
We estimate this by performing the integral up to $k = \Lambda = 1$,
then for $\tp=.05$
\begin{equation}
  \label{eq:deltaMF_kin}
  \delta E_{\text{kin}} \approx - 2 \bigl[ 1.9043  -
  \frac{\tp}{2} \ln (\Delta) \bigr]
  \Delta^2 + \ordo{\Delta^3},
\end{equation}
is the change in the kinetic energy per unit cell
due to $\Delta$.
The connection between the average magnetization 
$M =| <n_{j,\uparrow}> - <n_{j,\downarrow}>|$ and the mean field is
$\Delta = U M /2 $, where $U$ is the strength of the on-site Hubbard
interaction. 
The energy price one must pay per unit cell  
for having doubly occupied sites is
\begin{equation}
  \label{eq:deltaMF_U}
  \delta E_{U} = U M^2 = \frac{4}{U} \Delta^2.
\end{equation}
Because of the logarithm in Eq. (\ref{eq:deltaMF_kin}) a small
anti-ferromagnetic distortion is always favorable.
Assuming that $\Delta$ is small, the mean-field solution is:
\begin{equation}
  \label{eq:13}
  \Delta_{\text{MF}} \sim \exp 
  \Bigl(-\frac{1}{2} - \frac{2/U - 1.9043 }{\tp / 2} \Bigr),
\end{equation}
and hence $\Delta$ is exponentially suppressed unless $U$ is of
the order of $\Lambda$. 
Other variations of the mean-field in Eq.~(\ref{eq:Hkin0MF}) give
similar results.
Thus within the mean-field approximation,  
the anti-ferromagnetism found here is very weak unless the interaction
is strong.

\section{Conclusions}
\label{sec:conclusions}
In summary, we have shown that long-range Coulomb interactions in a clean graphene bilayer lead to a ground state 
that is magnetically polarized
with electron and hole pockets.  We have determined
the phase diagram of this model as a function of the coupling strength and doping, with a first order phase transition line between the  paramagnetic and ferromagnetic states. Around the critical line one expects hysteresis effects associated with the presence of phase coexistence and/or magnetic domains. 
We have also shown that on-site electron-electron interactions produce a staggering of the ferromagnetic order in the two planes and hence, c-axis antiferromagnetism.
The introduction of other terms in the Hamiltonian, such as trigonal
distortions, makes the phase diagram even richer, due to the creation of new energy scales. 
It is clear from our studies that graphene bilayers present an
electronic behavior that is  rather different from ordinary metals. The study of these systems becomes even more relevant given the recent developments in the fabrication and control of graphene multi-layers, and their possible application in nano-electronics. 

\begin{acknowledgments}
A.H.C.N. is supported through NSF grant DMR-0343790.
F.G. acknowledges funding from MEC(Spain) 
through grant FIS2005-05478-C02-01, 
and the European Union contract 12881(NEST).
\end{acknowledgments}

\appendix

\section{Unitary transformation}
\label{app:unitary}
The unitary transformation $\Mca_{\vp}$ that one needs to diagonalize
the Hamiltonian in Eq.~(\ref{eq:Hkin0}) can be written as $\Mca_{\vp} =
\Mca_1(\vp) \Mca_2 \Mca_3 (\vp)$, where
\begin{equation}
  \label{eq:UnitaryM1}
  \Mca_1(\vp) =
  \begin{pmatrix}
    1 & 0 & 0 & 0 \\
    0 & e^{-i \phi(\vp)} & 0 & 0\\
    0 & 0 & 1 & 0 \\
    0 & 0 & 0 & e^{i \phi(\vp)}
  \end{pmatrix}
\end{equation}
is a gauge transformation,
\begin{equation}
  \label{eq:UnitaryM2}
  \Mca_2 = \frac{1}{\sqrt{2}}
  \begin{pmatrix}
    1 & 0 & 1 & 0  \\
    0 & 1 & 0 & 1  \\
    1 & 0 & -1 & 0 \\
    0 & 1 & 0 & -1
  \end{pmatrix}
\end{equation}
forms symmetric/antisymmetric bands, and
\begin{equation}
  \label{eq:UnitaryM3}
  \Mca_3(\vp) \!\!=\!\! 
  \begin{pmatrix}
    \! \cos [\varphi(\vp)] & \sin [\varphi(\vp)] & 0 & 0  \\
    - \sin [\varphi(\vp)] & \cos [\varphi(\vp)] & 0 & 0  \\
    0 & 0 & \cos [\varphi(\vp)] & -\sin [\varphi(\vp)] \\
    0 & 0 & \sin [\varphi(\vp)] & \cos [\varphi(\vp)] \!
  \end{pmatrix} ,
\end{equation}
takes care of the final diagonalizing rotation. 
Choosing $\tan [2 \varphi(\vp)] = 2 p /\tp$ the rotated Hamiltonian 
$\Kca_{\text{diag}}(\vp) = \Mca^{\dag}_{\vp} \Kca(\vp) \Mca^{\,}_{\vp}$
 becomes diagonal:
\begin{multline}
  \label{eq:Kdiagonal}
    \Kca_{\text{diag}}(\vp) =
  \text{diag} \bigl[ -\tp/2 - E(p) , \:
    -\tp/2 + E(p) ,
    \\ \:
    \tp/2 + E(p) , \:
    \tp/2 - E(p) \bigr].
\end{multline}
Except for the labeling of the states ($1-4$), which is just a
permutation, this is the unitary transformation we need to diagonalize
the non-interacting problem.
\section{Exchange integral for Bloch electrons}
\label{app:Blochexchange}
Quite generally the Hartree-Fock energy consists of three terms.
A kinetic term, a direct charging term, and an exchange term. 
The exchange term is\cite{Herring66}
\begin{equation}
  \label{eq:Ex1}
  E_{\text{ex}} = \sum_{m,n}<m,n| V | n,m>,
\end{equation}
where $V$ is the interaction potential and
the sum is over occupied states. Expanding in Bloch states
$\psi_{\vk,\alpha}$ we get
\begin{equation}
  \label{eq:Ex2}
  E_{\text{ex}} = -\sum_{ \substack{\vk,\alpha,\sigma \\ \vk',\alpha'} }
  n_{\vk,\alpha,\sigma} n_{\vk',\alpha',\sigma} 
  J_{\vk,\vk'}^{\alpha, \alpha'},
\end{equation}
where for the unscreened Coulomb interaction in 2D
\begin{multline}
  \label{eq:Ex3}
  J_{\vk,\vk'}^{\alpha, \alpha'} = \frac{e^2}{2}
  \int d^2\vx \int d^2\vx'
\\ 
  \frac{\psi^{*}_{\vk,\alpha}(\vx) \psi^{\,}_{\vk',\alpha'}(\vx)
  \psi^{*}_{\vk',\alpha'}(\vx') \psi^{\,}_{\vk,\alpha}(\vx')}{|\vx -\vx'|}.
\end{multline}
Let us consider one plane only, then
we can write the Bloch states in the tight-binding approximation as
\begin{multline}
  \label{eq:Ex4}
  \psi^{\,}_{\vk,\alpha}(\vx) = \sum_{\vT_{\rA}}
  e^{i \vk \cdot \vT_{\rA} } a_{\vk,\alpha} w(\vx - \vT_{\rA} )
  \\  + \sum_{\vT_{\rB}}
   e^{i \vk \cdot \vT_{\rB}} b_{\vk,\alpha} w(\vx - \vT_{\rB} ) ,
\end{multline}
where $w(\vx)$ is the localized basis function, 
$\vT_{\rA}$ ($\vT_{\rB}$) are the lattice vectors of lattice A (B) and
$a_{\vk,\alpha}$ and $b_{\vk,\alpha}$ are the functions that
generate the Bloch state in question.
Neglecting the overlap of wave-functions on the A and B sites
we get approximately
\begin{multline}
  \label{eq:Ex5}
  \psi^{*}_{\vk,\alpha}(\vx) \psi^{\,}_{\vk',\alpha'}(\vx)
    \sim
   e^{i (\vk'- \vk) \cdot \vx } /S
\\ \times
   \bigl[ a^{*}_{\vk,\alpha} a^{\,}_{\vk',\alpha'} 
   \phi_{\rA}(\vx,\vk-\vk')
   + b^{*}_{\vk,\alpha} b^{\,}_{\vk',\alpha'} 
   \phi_{\rB}(\vx,\vk-\vk') \bigr].
\end{multline}
The functions $\phi_{\rA}$ and $\phi_{\rB}$ 
are periodic in the real-space 
lattice and can hence be expanded in components of 
harmonics of the reciprocal lattice $\{\vK \}$.
Keeping only the leading constant term coming from the $\vK = 0$ terms we
get $\phi = 1$, in which case
\begin{multline}
  \label{eq:Ex6}
  J_{\vk,\vk'}^{\alpha, \alpha'} = 
  \bigl[ a^{*}_{\vk,\alpha} a^{\,}_{\vk',\alpha'}
   + b^{*}_{\vk,\alpha} b^{\,}_{\vk',\alpha'}\bigr]
\\
   \times
   \bigl[ a^{*}_{\vk',\alpha'} a^{\,}_{\vk,\alpha}
   + b^{*}_{\vk',\alpha'} b^{\,}_{\vk,\alpha}\bigr] 
   \frac{2 \pi e^2}{S |\vk-\vk'|}.
\end{multline}
The corrections to this $\vK = 0$ term 
are down by at least a factor of
$|\vk-\vk'|^2$ as is explained in Ref.~[\onlinecite{Herring66}].
It is possible to include higher harmonics in the reciprocal lattice, 
but then the important
divergence near $\vk \approx \vk'$ is cut-off by $\vK$. Moreover, including 
$\vK \neq 0$ we should also include short-range (high energy) physics that
is not described by the continuum model used here.

Note also that the expression in Eq.~(\ref{eq:Ex6}) is just what one
get from a simple Fourier transform if one also includes the spinor
structure due to the two sub-lattices. We apply
this to the bilayer where
the Coulomb interaction can be written
\begin{multline}
  \label{eq:2planecoulomb}
  \Hex = \frac{1}{2}\int d^2\vx d^2\vy
  \Bigl\{
  V^{\rm D}(\vx - \vy) 
  \bigl[ \rho_1(\vx) \rho_1(\vy) + \rho_2(\vx) \rho_2(\vy) \bigr]
\\  +
  V^{\rm ND}(\vx - \vy) 
  \bigl[ \rho_1(\vx) \rho_2(\vy) + \rho_1(\vx) \rho_2(\vy) \bigr]
   \Bigr\}.
\end{multline}
We Fourier transform this and introduce the symmetric and
antisymmetric combinations
\begin{equation}
  \label{eq:2planerhoas}
  \rho_{\pm}(\vq) = \rho_1(\vq) \pm \rho_2(\vq).
\end{equation}
to write the interaction in a diagonal form
\begin{equation}
  \label{eq:2planecoulomb_sa}
  \Hex = \frac{1}{2 S} \sum'_{\vq} \sum_{\alpha=\pm}
  \rho_{\alpha}(\vq) V_{\alpha}(\vq) \rho_{\alpha}(-\vq),
\end{equation}
where $V_{\pm} = \frac{2 \pi e^2}{\eo q}(1 \pm e^{-q d})/2$. The prime on
the $\vq$-sum denotes that the $\vq =0$ term should be excluded
since it is canceled by the positive (jellium) background.
In terms of the operators that diagonalizes the kinetic terms the
density operators can be written as
\begin{equation}
  \label{eq:density}
  \rho_{\pm}(\vq) = \sum_{\vp} \Phi_{\vp + \vq}^{\dag} 
  \chi^{\pm}(\vp + \vq,\vp) \Phi_{\vp},
\end{equation}
where
\begin{equation}
  \label{eq:chidef}
  \chi^{\pm}(\vp + \vq,\vp) \equiv \Mca_{\vp + \vq}^{\dag}
  \begin{pmatrix}
    1 & 0 & 0 & 0 \\
    0 & 1 & 0 & 0 \\
    0 & 0 & \pm 1 & 0 \\
    0 & 0 & 0 & \pm 1
  \end{pmatrix}
  \Mca_{\vp},
\end{equation}
and $\Mca_{\vp}$ is given in Appendix~\ref{app:unitary}.
Using this one can easily 
generalize Eq.~(\ref{eq:Ex2}) and (\ref{eq:Ex6})
to arrive at Eq.~(\ref{eq:Hexvar}).

\section{Exchange integral}
\label{app:exchange_integral}
Using $V^{\rD}(\vq) = 2\pi g / q$
and $V^{\rND}(\vq) = 2\pi g e^{-q d} / q$ one quite generally get
that the change in the exchange energy can be written as
\begin{multline}
  \label{eq:2plane_deltaE_ex1}
  <\frac{ \Delta \Hex}{S} > =
\\ - \frac{1}{2} 
  \int \frac{d \bm p}{(2 \pi)^2} \frac{d \bm p'}{(2 \pi)^2}
  \Bigr\{
  2 \Theta(Q-p) \bigl[ K_{1}^{\rD} V^{\rD} + K_{1}^{\rND} V^{\rND} \bigr]
\\  
  + \Theta(Q-p) \Theta(Q-p') \bigl[ K_{2}^{\rD} V^{\rD} + K_{2}^{\rND} V^{\rND} \bigr]
  \Bigr\} \, .
\end{multline}
The $K$'s are sums of certain components of the matrices 
$\xi^{\alpha}_{ij}(\vp',\vp) = 
\chi^{\alpha}_{ij}(\vp',\vp)
\chi^{\alpha}_{ji}(\vp, \vp')
$. Note that the elements of $\xi$ are all greater or equal to zero.
\subsection{Half-filling}
At half-filling the trial state is characterized by the single
variational parameter $Q$. 
It is straightforward, albeit tedious, to perform the matrix
multiplications and extract the $K$'s.
The results for an electron \textit{and} a hole pocket of size $Q$ is:
\begin{equation}
  \label{eq:twoconeK1D}
  K_1^{\rD} = - \frac{p}{E(p)} \frac{p'}{E(p')} \cos(\phi -\phi') ,
\end{equation}
\begin{equation}
  \label{eq:twoconeK1ND}
  K_1^{\rND} = \frac{t}{2 E(p')} \Bigl\{ \bigl[1 -\frac{t}{E(p)} \bigr]
  - \bigl[1 + \frac{t}{E(p)} \bigr] \cos 2(\phi -\phi') ,
\end{equation}
\begin{equation}
  \label{eq:twoconeK2D}
  K_2^{\rD} = \frac{1}{2} \frac{p}{E(p)} \frac{p'}{E(p')} \cos(\phi -\phi')
  +\frac{1}{2} \bigl[ 1+ \frac{t}{E(p)} \frac{t}{E(p')} \bigr] ,
\end{equation}
\begin{multline}
  \label{eq:twoconeK2ND}
  K_2^{\rND} = \frac{1}{2} \frac{p}{E(p)} \frac{p'}{E(p')} \cos(\phi -\phi')
\\
  +\frac{1}{4} \Bigl\{ \bigl[1- \frac{t}{E(p)}\bigr] \bigl[1 -
  \frac{t}{E(p')}\bigr] \\ +
\bigl[1+ \frac{t}{E(p)}\bigr] \bigl[1+\frac{t}{E(p')}\bigr] \cos
  2(\phi -\phi')
 \Bigr\}.
\end{multline}
In this appendix we temporarily relabel $\tp/2 \rightarrow t$ to avoid an
excessive amount of $2$'s in the equations.
We can write
\begin{equation}
  \label{eq:2plane_deltaE_ex2}
  <\frac{\Delta \Hex}{S}> = \frac{g \vf \Lambda^3 }{4 \pi^2} 
  \bigl[ C_1^{\rD} + C_1^{\rND} + C_2^{\rD} + C_2^{\rND} \bigr],
\end{equation}
where we are measuring all parameters (i.e. $Q$ and $t$) in units of
the cut-off $\Lambda$.
The $C$'s are given by the integrals
\begin{widetext}
\begin{eqnarray}
  C_1^{\rD} &=& 2 Q^3 
  \int_0^1 x dx \int_0^1 y dy 
  \int_0^{\pi}d\phi 
  \frac{x}{\sqrt{t^2+(Q x)^2}}
  \frac{y}{\sqrt{y^2+t^2}} 
  \frac{\cos(\phi)}{|\vy - Q \vx|},
\nonumber \\
  C_1^{\rND} &=& - Q^2 
  \int_0^1 x dx \int_0^1 y dy 
  \int_0^{\pi}d\phi 
  \frac{t}{\sqrt{t^2+y^2}}
  \frac{e^{-d \Lambda |\vy - Q \vx|}}{|\vy - Q \vx|}
  \Bigl\{
  \bigl[1 -\frac{t}{\sqrt{t^2+(Q x)^2}}\bigr]
  -\bigl[1 +\frac{t}{\sqrt{t^2 +(Q x)^2}}\bigr]\cos(2\phi)
  \Bigl\},
\nonumber \\
  C_2^{\rD} &=&  -\frac{Q^3}{2} 
  \int_0^1 x dx \int_0^1 y dy 
  \int_0^{\pi}d\phi \Bigl\{ Q^2
  \bigl[
  \frac{x}{\sqrt{t^2 +(Q x)^2}}
  \frac{y}{\sqrt{t^2 +(Q y)^2}}
  \bigr] \cos(\phi)
\nonumber \\ & &
+ \bigl[1+\frac{t}{\sqrt{t^2 +(Q y)^2}} \frac{t}{\sqrt{t^2 +(Q x)^2}}\bigr]
\Bigr\}
\frac{1}{|\vy - \vx|},
\nonumber \\
  C_2^{\rND} &=& - \frac{Q^3}{4} 
  \int_0^1 x dx \int_0^1 y dy 
  \int_0^{\pi}d\phi 
  \frac{e^{-d \Lambda Q |\vy -\vx|}}{|\vy - \vx|}
  \biggl(
  2 Q^2
  \bigl[
  \frac{x}{\sqrt{t^2 +(Q x)^2}}
  \frac{y}{\sqrt{t^2 +(Q y)^2}}
  \bigr] \cos(\phi) 
\nonumber \\ &+ &
  \bigl[1-\frac{t}{\sqrt{t^2 +(Q x)^2}}\bigr]
  \bigl[1-\frac{t}{\sqrt{t^2 +(Q y)^2}}\bigr]
+
  \bigl[1+\frac{t}{\sqrt{t^2 +(Q x)^2}}\bigr]
  \bigl[1+\frac{t}{\sqrt{t^2 +(Q y)^2}}\bigr]
  \cos(2\phi)
  \biggr).
\end{eqnarray}
\end{widetext}
From this one can extract the leading and sub-leading
terms in an expansion in
powers of $Q$, the result is given in Eq.~(\ref{eq:effectiveenergy1}).
Some useful expressions for performing the expansion are provided in Appendix
\ref{app:Coulomb_integrals}.
\subsection{Doped case}
The calculation for the doped system proceeds exactly as in the previous case
but we must allow for the electron and hole pockets to have
different size.
For each electron (or hole) pocket of size $Q_{e}$ ($Q_{h}$) 
there is a contribution like that in Eq. (\ref{eq:2plane_deltaE_ex1}).
The $K$'s are half of those in 
Eq.~(\ref{eq:twoconeK1ND}), (\ref{eq:twoconeK2D}) and (\ref{eq:twoconeK2ND}).
But $K^{\rD}_1$ is different for holes and electrons:
\begin{equation}
  \label{eq:KD1eh}
  K^{\rD}_{1e/h} = \pm \frac{1}{2} - \frac{p \, p'}{2 E(p) E(p')} \cos(\theta).
\end{equation}
The contributions coming from the $\pm1/2$ are easy to obtain and can
be encoded in a new contribution $C_{\text{new}}$ in 
Eq.~(\ref{eq:2plane_deltaE_ex2}), where
\begin{equation}
  \label{eq:C1new}
  C_{\text{new}} = - Q^2 R_0 (Q), 
\end{equation}
and $R_0$ is given in Appendix \ref{app:Coulomb_integrals}.

\section{Coulomb integrals}
\label{app:Coulomb_integrals}
Let us define
\begin{equation}
  \label{eq:Rdef}
  R_n (Q) = \int_0^1 dx \int_0^{1} dy 
  \int_0^{\pi}d\phi   \frac{x y \cos(n\phi)}{\sqrt{Q^2 x^2 + y^2 -2 Q x y \cos(\phi)}}.
\end{equation}
Then one can show that up to $\ordo{Q^{10}}$
\begin{subequations}
  \begin{eqnarray}
    R_0 (Q) &= & \frac{\pi}{2} \Bigl[
    1 - \frac{Q^2}{8} - \frac{Q^4}{64} - \frac{5 \, Q^6}{1024} - \frac{35
   \, Q^8}{16384} \Bigr] ,
\nonumber \\
    R_1 (Q) &=&
    - \frac{\pi}{6} Q \ln(Q) 
  + \frac{ \pi }{3} \bigl[ \ln(2)- \frac{1}{12}\bigr] \, Q
\nonumber \\
    &-& \frac{\pi}{2} \Bigl[ \frac{3 \, Q^3}{80} 
    + \frac{15 \, Q^5}{1792} + \frac{175 \, Q^7}{55296} 
    + \frac{2205 \, Q^9}{1441792} \Bigr] ,
\nonumber \\
    R_2 (Q) &= & \frac{2 \pi}{9} Q 
    -
    \frac{\pi}{2} \Bigl[
    \frac{3\, Q^2}{16} + \frac{5 \, Q^4}{288} + \frac{21 \, Q^6}{4096} 
+ \frac{9 \, Q^8}{4096} \Bigr]. 
\nonumber
\end{eqnarray}
\end{subequations}
Moreover $R_0 (1) = 4/3$, $R_0 (1) = 2 (2 \mathcal{C} -1)/3$ and
$R_2 (1) = 4 (3 \pi - 7)/27$. $\mathcal{C} \approx 0.91596$ is the Catalan
constant.

\section{Approximate two-band models}
\label{app:effective}

There are two reasons for constructing approximate two-band
models. Firstly, on physical grounds the high-energy bands should not
be very important for the low-energy properties of the system.
Secondly, it is much easier to work with $2\times2$ matrices instead
of $4\times4$ matrices.
In this appendix we derive the low-energy effective model by doing
degenerate second order perturbation theory. The quality of the
expansion is good as long as $\vf p \ll \tp$.

\subsection{Simple low-energy model}
If we transform the Hamiltonian matrix in Eq. (\ref{eq:Hkin0}) 
by taking the symmetric and anti-symmetric combinations of the first
and third rows (and columns) we can write $\Hkin = \Hca_0 + \Hca_1$, where
\begin{equation}
  \Hca_0 = 
  \begin{pmatrix}
    -\tp & 0 & 0   & 0 \\
    0    & 0 & 0   & 0 \\
    0    & 0 & \tp & 0 \\
    0    & 0 & 0   & 0
  \end{pmatrix},
\end{equation}
\begin{equation}
  \Hca_1
  = \frac{1}{\sqrt{2}}
  \begin{pmatrix}
    0 & p e^{i \phi(\bm p)} & 0 & p e^{-i \phi(\bm p)} \\
    p e^{-i \phi(\bm p)} & 0 & p e^{-i \phi(\bm p)} & 0\\
    0 & p e^{i \phi(\bm p)} & 0 & -p e^{-i \phi(\bm p)} \\
    p e^{i \phi(\bm p)} & 0 & -p e^{i \phi(\bm p)} & 0
  \end{pmatrix}.
\end{equation}
Performing second order perturbation theory one finds an effective
Hamiltonian matrix, $\Hca_{\text{eff}} = - \Hca^{\dag}_1 (1/\Hca_0) \Hca_1$,
where:
\begin{equation}
  \Hca_{\text{eff}}
  = \frac{p^2}{\tp}
  \begin{pmatrix}
    0 & 0 & 0 & 0 \\
    0 & 0 & 0 & e^{-2 i \phi(\bm p)}\\
    0 & 0 & 0 & 0 \\
    0 & e^{2 i \phi(\bm p)} & 0 & 0
  \end{pmatrix}.
\end{equation}
The low-energy spinors are then given by:
\begin{equation}
  \label{eq:effectivespinors}
  \Psi_{\pm}(\bm p) = \frac{1}{\sqrt{2}}
  \begin{pmatrix}  0 \\ e^{-i \phi(\bm p)} \\ 0 \\ \pm e^{i \phi(\bm p)} 
    \end{pmatrix},
\end{equation}
and the corresponding energies are $\pm p^2 / \tp$.
If we add the contribution from $\gamma_3$ (which is already diagonal in
this basis) and only keep the B-atom components 
we immediately arrive at Eq. (\ref{eq:Ewarping}).

In fact it is easier to see the existence of the exchange instability
in this basis. Working in this subspace 
we again get an expression like that in 
Eq.~(\ref{eq:2plane_deltaE_ex1}) with 
$K_1^{\text{D}}=0$, $K_1^{\text{ND}} = - \cos(2\phi)$, $K^{\text{D}}_2 = 1$ 
and $K^{\text{ND}}_2 = \cos(2 \phi)$. 
If one further neglects the difference between $V^{\text{D}}$ and
$V^{\text{ND}}$ the resulting change in the exchange energy can be
expressed with the help of the functions defined in Appendix
\ref{app:Coulomb_integrals}.
Explicitly the leading term in the exchange energy is 
$\propto -Q^3 \bigl[ R_0(1) + R_2(1) \bigr] + 2 Q^2
R_2(Q) \sim - 8 Q^3 /27$ in agreement with the result in
Eq.~(\ref{eq:effectiveenergy1}).

\subsection{More general low-energy model}
A more general Hamiltonian model for the low energy physics is given
by \cite{Wallace47}: 
\begin{equation}
  \label{eq:1}
  \Hca_0 = 
  \begin{pmatrix}
    0 & \vf p e^{i \phi} & \tp & v_4  p e^{-i \phi}
    \\
    \vf  p e^{-i \phi} & 0 & v_4  p e^{-i \phi} & v_3  p
    e^{i \phi} \\
    \tp & v_4  p e^{i \phi} & 0 & \vf p  e^{-i \phi} \\
    v_4  p e^{i \phi} & v_3  p e^{-i \phi} & \vf p e^{i \phi} & 0
  \end{pmatrix}.
\end{equation}
One can perform a unitary transformation so that
$\Hca_2 = \Mca^{\dag} \Hca_0 \Mca$, 
where $\Mca = \Mca_1(\vp) \Mca_2$ and $\Mca_1(\vp)$
and $\Mca_2$ are given in Eq.~(\ref{eq:UnitaryM1}) and
(\ref{eq:UnitaryM2}).
We may then separate the transformed Hamiltonian into three parts, 
$\Hca_2 = \Kca_0 + \Kca_1 +\Kca_2$, with:
\begin{eqnarray}
  \label{eq:Kca0}
  \Kca_{0} &=&
    \begin{pmatrix}
    \tp & 0 & 0 & 0  \\
    0 & 0 & 0 & 0 \\
    0 & 0 & -\tp & 0 \\  
    0 & 0 & 0 & 0
  \end{pmatrix},
%
\\
%
  \label{eq:Kca1}
  \Kca_{1} &=&
    \begin{pmatrix}
    0 & 0 & 0 & 0  \\
    0 & v_3  p \cos(3 \phi) &  
    0 & -i v_3  p \sin(3 \phi) \\
    0 & 0 & 0 & 0 \\  
    0 & i v_3  p \sin( 3 \phi) & 0 & - v_3  p \cos(3 \phi)
  \end{pmatrix},
%
\\
%
  \label{eq:Kca2}
  \Kca_{2} &=&
    \begin{pmatrix}
    0 & (\vf + v_4 ) p &  0 & 0   \\
    (\vf + v_4 ) p & 0 &  0 & 0 \\
    0 & 0 & 0 & (\vf - v_4 ) p  \\  
    0 & 0 & (\vf - v_4 ) p & 0
  \end{pmatrix}.
\end{eqnarray}
With this decomposition it is easy to find
the approximate eigenstates and eigenvalues for $\vf p \ll \tp$.
\begin{widetext}
For the high-energy states one can use the simple non-degenerate 
perturbation theory. The eigenvalues are given by
$E_3 =  \tp + (\vf +  v_4 )^2 p^2 /  \tp$ and
$E_4 = - \tp - (\vf -  v_4 )^2 p^2 /  \tp$. It is also straightforward
to obtain the corresponding states.
For the low-energy sector the second order perturbation result (from
two $\Kca_2$ and one $\Kca_0$) can
give a term which is of the same order as that
of $\Kca_1$. Thus, we must use degenerate perturbation theory.
The usual manipulations then given the Hamiltonian matrix in the low
energy subspace as 
$\Kca_{\text{low}} = \Kca_1 - \Kca_2 \Pca_1 (1/\Kca_0) \Pca_1 \Kca_2$,
where $\Pca_1$ is the projection out of the low-energy subspace,
explicitly
\begin{equation}
  \Kca_{\text{low}} =  
  \label{eq:lowe_matirx}
    \begin{pmatrix}
    0 & 0 & 0 & 0  \\
    0 &  v_3  p \cos(3 \phi) &  
    0 &  -i v_3  p \sin(3 \phi) \\
    0 & 0 & 0 & 0 \\  
    0 &  i v_3  p \sin( 3 \phi) & 0 & - v_3  p \cos(3 \phi)
  \end{pmatrix}
+
  \frac{p^2}{ \tp}
  \begin{pmatrix}
    0 & 0 & 0 & 0  \\
    0 & -(\vf +  v_4)^2 & 0 & 0 \\
    0 & 0 & 0 & 0 \\  
    0 & 0 & 0 & (\vf -  v_4)^2
   \end{pmatrix},
\end{equation}
and the corresponding eigenvalues are
  \begin{equation}
    \label{eq:lowebands}
  -\frac{2 \vf v_4 p^2}{ \tp}
  \pm \sqrt{( v_3 p)^2 
    + \Bigl[ \frac{(\vf^2 + v_{4}^2) p^2}{ \tp} \Bigr]^2
  - 2 ( v_3 p ) \Bigl[ \frac{(\vf^2 + v_{4}^2) p^2}{ \tp} \Bigr] 
    \cos(3 \phi)
  }    .
  \end{equation}
\end{widetext}

\bibliography{graphite13.bib}

\newcommand{\npb}{Nucl. Phys.}\newcommand{\adv}{Adv.
  Phys.}\newcommand{\epl}{Europhys. Lett.}
\begin{thebibliography}{25}
\expandafter\ifx\csname natexlab\endcsname\relax\def\natexlab#1{#1}\fi
\expandafter\ifx\csname bibnamefont\endcsname\relax
  \def\bibnamefont#1{#1}\fi
\expandafter\ifx\csname bibfnamefont\endcsname\relax
  \def\bibfnamefont#1{#1}\fi
\expandafter\ifx\csname citenamefont\endcsname\relax
  \def\citenamefont#1{#1}\fi
\expandafter\ifx\csname url\endcsname\relax
  \def\url#1{\texttt{#1}}\fi
\expandafter\ifx\csname urlprefix\endcsname\relax\def\urlprefix{URL }\fi
\providecommand{\bibinfo}[2]{#2}
\providecommand{\eprint}[2][]{\url{#2}}

\bibitem[{\citenamefont{Novoselov et~al.}(2005)\citenamefont{Novoselov, Geim,
  Morozov, Jiang, Katsnelson, Grigorieva, Dubonos, and Firsov}}]{Novolelov2005}
\bibinfo{author}{\bibfnamefont{K.~S.} \bibnamefont{Novoselov}},
  \bibinfo{author}{\bibfnamefont{A.~K.} \bibnamefont{Geim}},
  \bibinfo{author}{\bibfnamefont{S.~V.} \bibnamefont{Morozov}},
  \bibinfo{author}{\bibfnamefont{D.}~\bibnamefont{Jiang}},
  \bibinfo{author}{\bibfnamefont{M.~I.} \bibnamefont{Katsnelson}},
  \bibinfo{author}{\bibfnamefont{I.~V.} \bibnamefont{Grigorieva}},
  \bibinfo{author}{\bibfnamefont{S.~V.} \bibnamefont{Dubonos}},
  \bibnamefont{and} \bibinfo{author}{\bibfnamefont{A.~A.}
  \bibnamefont{Firsov}}, \bibinfo{journal}{Nature}
  \textbf{\bibinfo{volume}{438}}, \bibinfo{pages}{197} (\bibinfo{year}{2005}).

\bibitem[{\citenamefont{Zhang et~al.}(2005)\citenamefont{Zhang, Tan, Stormer,
  and Kim}}]{Zhang2005}
\bibinfo{author}{\bibfnamefont{Y.}~\bibnamefont{Zhang}},
  \bibinfo{author}{\bibfnamefont{Y.-W.} \bibnamefont{Tan}},
  \bibinfo{author}{\bibfnamefont{H.~L.} \bibnamefont{Stormer}},
  \bibnamefont{and} \bibinfo{author}{\bibfnamefont{P.}~\bibnamefont{Kim}},
  \bibinfo{journal}{Nature} \textbf{\bibinfo{volume}{438}},
  \bibinfo{pages}{201} (\bibinfo{year}{2005}).

\bibitem[{\citenamefont{Wallace}(1947)}]{Wallace47}
\bibinfo{author}{\bibfnamefont{P.~R.} \bibnamefont{Wallace}},
  \bibinfo{journal}{Phys. Rev.} \textbf{\bibinfo{volume}{71}},
  \bibinfo{pages}{622} (\bibinfo{year}{1947}).

\bibitem[{\citenamefont{Peres et~al.}()\citenamefont{Peres, Guinea, and {Castro
  Neto}}}]{nuno_curto}
\bibinfo{author}{\bibfnamefont{N.~M.~R.} \bibnamefont{Peres}},
  \bibinfo{author}{\bibfnamefont{F.}~\bibnamefont{Guinea}}, \bibnamefont{and}
  \bibinfo{author}{\bibfnamefont{A.~H.} \bibnamefont{{Castro Neto}}},
  \bibinfo{note}{cond-mat/0506709}.

\bibitem[{\citenamefont{Peres et~al.}(2006)\citenamefont{Peres, Guinea, and
  {Castro Neto}}}]{nuno2006_long}
\bibinfo{author}{\bibfnamefont{N.~M.~R.} \bibnamefont{Peres}},
  \bibinfo{author}{\bibfnamefont{F.}~\bibnamefont{Guinea}}, \bibnamefont{and}
  \bibinfo{author}{\bibfnamefont{A.~H.} \bibnamefont{{Castro Neto}}},
  \bibinfo{journal}{Phys. Rev. B.} \textbf{\bibinfo{volume}{73}},
  \bibinfo{pages}{125411} (\bibinfo{year}{2006}).

\bibitem[{\citenamefont{Brandt et~al.}(1988)\citenamefont{Brandt, Chudinov, and
  Ponomarev}}]{BCP88}
\bibinfo{author}{\bibfnamefont{N.~B.} \bibnamefont{Brandt}},
  \bibinfo{author}{\bibfnamefont{S.~M.} \bibnamefont{Chudinov}},
  \bibnamefont{and} \bibinfo{author}{\bibfnamefont{Y.~G.}
  \bibnamefont{Ponomarev}}, in \emph{\bibinfo{booktitle}{Modern Problems in
  Condensed Matter Sciences}}, edited by \bibinfo{editor}{\bibfnamefont{V.~M.}
  \bibnamefont{Agranovich}} \bibnamefont{and}
  \bibinfo{editor}{\bibfnamefont{A.~A.} \bibnamefont{Maradudin}}
  (\bibinfo{publisher}{North Holland (Amsterdam)}, \bibinfo{year}{1988}), vol.
  \bibinfo{volume}{20.1}.

\bibitem[{\citenamefont{Rydberg et~al.}(2003)\citenamefont{Rydberg, Dion,
  Jacobson, Schroder, Hyldgaard, Simak, Langreth, and Lundqvist}}]{Rydberg2003}
\bibinfo{author}{\bibfnamefont{H.}~\bibnamefont{Rydberg}},
  \bibinfo{author}{\bibfnamefont{M.}~\bibnamefont{Dion}},
  \bibinfo{author}{\bibfnamefont{N.}~\bibnamefont{Jacobson}},
  \bibinfo{author}{\bibfnamefont{E.}~\bibnamefont{Schroder}},
  \bibinfo{author}{\bibfnamefont{P.}~\bibnamefont{Hyldgaard}},
  \bibinfo{author}{\bibfnamefont{S.}~\bibnamefont{Simak}},
  \bibinfo{author}{\bibfnamefont{D.}~\bibnamefont{Langreth}}, \bibnamefont{and}
  \bibinfo{author}{\bibfnamefont{B.}~\bibnamefont{Lundqvist}},
  \bibinfo{journal}{Phys. Rev. Lett.} \textbf{\bibinfo{volume}{91}},
  \bibinfo{pages}{126402} (\bibinfo{year}{2003}).

\bibitem[{\citenamefont{Esquinazi et~al.}(2003)\citenamefont{Esquinazi,
  Spemann, H\"ohne, Setzer, Han, and Butz}}]{esquinazi}
\bibinfo{author}{\bibfnamefont{P.}~\bibnamefont{Esquinazi}},
  \bibinfo{author}{\bibfnamefont{D.}~\bibnamefont{Spemann}},
  \bibinfo{author}{\bibfnamefont{R.}~\bibnamefont{H\"ohne}},
  \bibinfo{author}{\bibfnamefont{A.}~\bibnamefont{Setzer}},
  \bibinfo{author}{\bibfnamefont{K.-H.} \bibnamefont{Han}}, \bibnamefont{and}
  \bibinfo{author}{\bibfnamefont{T.}~\bibnamefont{Butz}},
  \bibinfo{journal}{Phys. Rev. Lett.} \textbf{\bibinfo{volume}{91}},
  \bibinfo{pages}{227201} (\bibinfo{year}{2003}).

\bibitem[{\citenamefont{Vozmediano et~al.}(2005)\citenamefont{Vozmediano,
  L\'opez-Sancho, Stauber, and Guinea}}]{Vetal05}
\bibinfo{author}{\bibfnamefont{M.~A.~H.} \bibnamefont{Vozmediano}},
  \bibinfo{author}{\bibfnamefont{M.~P.} \bibnamefont{L\'opez-Sancho}},
  \bibinfo{author}{\bibfnamefont{T.}~\bibnamefont{Stauber}}, \bibnamefont{and}
  \bibinfo{author}{\bibfnamefont{F.}~\bibnamefont{Guinea}},
  \bibinfo{journal}{Phys. Rev. B} \textbf{\bibinfo{volume}{72}},
  \bibinfo{pages}{155121} (\bibinfo{year}{2005}).

\bibitem[{\citenamefont{Peres et~al.}(2005)\citenamefont{Peres, Guinea, and
  {Castro Neto}}}]{Peres2005_ferro}
\bibinfo{author}{\bibfnamefont{N.~M.~R.} \bibnamefont{Peres}},
  \bibinfo{author}{\bibfnamefont{F.}~\bibnamefont{Guinea}}, \bibnamefont{and}
  \bibinfo{author}{\bibfnamefont{A.~H.} \bibnamefont{{Castro Neto}}},
  \bibinfo{journal}{Phys. Rev. B} \textbf{\bibinfo{volume}{72}},
  \bibinfo{pages}{174406} (\bibinfo{year}{2005}).

\bibitem[{\citenamefont{Bloch}(1929)}]{Bloch}
\bibinfo{author}{\bibfnamefont{F.}~\bibnamefont{Bloch}}, \bibinfo{journal}{Z.
  Physik} \textbf{\bibinfo{volume}{57}}, \bibinfo{pages}{549}
  (\bibinfo{year}{1929}).

\bibitem[{\citenamefont{Berger et~al.}(2004)\citenamefont{Berger, Song, Li, Li,
  Ogbazghi, Feng, Dai, Marchenkov, Conrad, First et~al.}}]{berger04}
\bibinfo{author}{\bibfnamefont{C.}~\bibnamefont{Berger}},
  \bibinfo{author}{\bibfnamefont{Z.}~\bibnamefont{Song}},
  \bibinfo{author}{\bibfnamefont{T.}~\bibnamefont{Li}},
  \bibinfo{author}{\bibfnamefont{X.}~\bibnamefont{Li}},
  \bibinfo{author}{\bibfnamefont{A.~Y.} \bibnamefont{Ogbazghi}},
  \bibinfo{author}{\bibfnamefont{R.}~\bibnamefont{Feng}},
  \bibinfo{author}{\bibfnamefont{Z.}~\bibnamefont{Dai}},
  \bibinfo{author}{\bibfnamefont{A.~N.} \bibnamefont{Marchenkov}},
  \bibinfo{author}{\bibfnamefont{E.~H.} \bibnamefont{Conrad}},
  \bibinfo{author}{\bibfnamefont{P.~N.} \bibnamefont{First}},
  \bibnamefont{et~al.}, \bibinfo{journal}{J. Phys. Chem.}
  \textbf{\bibinfo{volume}{108}}, \bibinfo{pages}{19912}
  (\bibinfo{year}{2004}).

\bibitem[{\citenamefont{Guinea et~al.}(2000)\citenamefont{Guinea,
  G\'omez-Santos, and Arovas}}]{GGA00}
\bibinfo{author}{\bibfnamefont{F.}~\bibnamefont{Guinea}},
  \bibinfo{author}{\bibfnamefont{G.}~\bibnamefont{G\'omez-Santos}},
  \bibnamefont{and} \bibinfo{author}{\bibfnamefont{D.}~\bibnamefont{Arovas}},
  \bibinfo{journal}{Phys. Rev. B} \textbf{\bibinfo{volume}{62}},
  \bibinfo{pages}{391} (\bibinfo{year}{2000}).

\bibitem[{\citenamefont{Du et~al.}(2005)\citenamefont{Du, Tsai, Maslov, and
  Hebard}}]{Duetal04}
\bibinfo{author}{\bibfnamefont{X.}~\bibnamefont{Du}},
  \bibinfo{author}{\bibfnamefont{S.-W.} \bibnamefont{Tsai}},
  \bibinfo{author}{\bibfnamefont{D.~L.} \bibnamefont{Maslov}},
  \bibnamefont{and} \bibinfo{author}{\bibfnamefont{A.~F.}
  \bibnamefont{Hebard}}, \bibinfo{journal}{Phys. Rev. Lett.}
  \textbf{\bibinfo{volume}{94}}, \bibinfo{pages}{166601}
  (\bibinfo{year}{2005}).

\bibitem[{\citenamefont{McClure}(1957)}]{M57}
\bibinfo{author}{\bibfnamefont{J.~W.} \bibnamefont{McClure}},
  \bibinfo{journal}{Phys. Rev.} \textbf{\bibinfo{volume}{108}},
  \bibinfo{pages}{612} (\bibinfo{year}{1957}).

\bibitem[{\citenamefont{Slonczewski and Weiss}(1958)}]{SW58}
\bibinfo{author}{\bibfnamefont{J.~C.} \bibnamefont{Slonczewski}}
  \bibnamefont{and} \bibinfo{author}{\bibfnamefont{P.~R.} \bibnamefont{Weiss}},
  \bibinfo{journal}{Phys. Rev.} \textbf{\bibinfo{volume}{109}},
  \bibinfo{pages}{272} (\bibinfo{year}{1958}).

\bibitem[{\citenamefont{McCann and Fal'ko}(2006)}]{Falko2006}
\bibinfo{author}{\bibfnamefont{E.}~\bibnamefont{McCann}} \bibnamefont{and}
  \bibinfo{author}{\bibfnamefont{V.~I.} \bibnamefont{Fal'ko}},
  \bibinfo{journal}{Phys. Rev. Lett.} \textbf{\bibinfo{volume}{96}},
  \bibinfo{pages}{086805} (\bibinfo{year}{2006}).

\bibitem[{\citenamefont{Peres et~al.}(2004)\citenamefont{Peres, Ara\`ujo, and
  Bozi}}]{Peres03}
\bibinfo{author}{\bibfnamefont{N.~M.~R.} \bibnamefont{Peres}},
  \bibinfo{author}{\bibfnamefont{M.~A.~N.} \bibnamefont{Ara\`ujo}},
  \bibnamefont{and} \bibinfo{author}{\bibfnamefont{D.}~\bibnamefont{Bozi}},
  \bibinfo{journal}{Phys. Rev. B} \textbf{\bibinfo{volume}{70}},
  \bibinfo{pages}{195122} (\bibinfo{year}{2004}).

\bibitem[{\citenamefont{Gonz\'alez et~al.}(1997)\citenamefont{Gonz\'alez,
  Guinea, and Vozmediano}}]{GGV97b}
\bibinfo{author}{\bibfnamefont{J.}~\bibnamefont{Gonz\'alez}},
  \bibinfo{author}{\bibfnamefont{F.}~\bibnamefont{Guinea}}, \bibnamefont{and}
  \bibinfo{author}{\bibfnamefont{M.~A.~H.} \bibnamefont{Vozmediano}},
  \bibinfo{journal}{Nucl. Phys. B} \textbf{\bibinfo{volume}{485 [FS]}},
  \bibinfo{pages}{694} (\bibinfo{year}{1997}).

\bibitem[{\citenamefont{Gonz\'alez et~al.}(1994)\citenamefont{Gonz\'alez,
  Guinea, and Vozmediano}}]{GGV94b}
\bibinfo{author}{\bibfnamefont{J.}~\bibnamefont{Gonz\'alez}},
  \bibinfo{author}{\bibfnamefont{F.}~\bibnamefont{Guinea}}, \bibnamefont{and}
  \bibinfo{author}{\bibfnamefont{M.~A.~H.} \bibnamefont{Vozmediano}},
  \bibinfo{journal}{Nucl. Phys. B} \textbf{\bibinfo{volume}{424 [FS]}},
  \bibinfo{pages}{595} (\bibinfo{year}{1994}).

\bibitem[{\citenamefont{Gonz\'alez et~al.}(1996)\citenamefont{Gonz\'alez,
  Guinea, and Vozmediano}}]{GGV96}
\bibinfo{author}{\bibfnamefont{J.}~\bibnamefont{Gonz\'alez}},
  \bibinfo{author}{\bibfnamefont{F.}~\bibnamefont{Guinea}}, \bibnamefont{and}
  \bibinfo{author}{\bibfnamefont{M.~A.~H.} \bibnamefont{Vozmediano}},
  \bibinfo{journal}{Phys. Rev. Lett.} \textbf{\bibinfo{volume}{77}},
  \bibinfo{pages}{3589} (\bibinfo{year}{1996}).

\bibitem[{\citenamefont{Mahan}(2000)}]{mahan}
\bibinfo{author}{\bibfnamefont{G.~D.} \bibnamefont{Mahan}},
  \emph{\bibinfo{title}{Many-Particle Physics}} (\bibinfo{publisher}{Plenum},
  \bibinfo{year}{2000}).

\bibitem[{\citenamefont{Canel et~al.}(1972)\citenamefont{Canel, Matthews, and
  Zia}}]{Canel1972}
\bibinfo{author}{\bibfnamefont{E.}~\bibnamefont{Canel}},
  \bibinfo{author}{\bibfnamefont{M.~P.} \bibnamefont{Matthews}},
  \bibnamefont{and} \bibinfo{author}{\bibfnamefont{R.~K.~P.}
  \bibnamefont{Zia}}, \bibinfo{journal}{Phys. kondens. Materie}
  \textbf{\bibinfo{volume}{15}}, \bibinfo{pages}{191} (\bibinfo{year}{1972}).

\bibitem[{\citenamefont{Fetter}(1973)}]{Fetter1973}
\bibinfo{author}{\bibfnamefont{A.~L.} \bibnamefont{Fetter}},
  \bibinfo{journal}{Ann. Phys.} \textbf{\bibinfo{volume}{81}},
  \bibinfo{pages}{367} (\bibinfo{year}{1973}).

\bibitem[{\citenamefont{Herring}(1966)}]{Herring66}
\bibinfo{author}{\bibfnamefont{C.}~\bibnamefont{Herring}},
  \emph{\bibinfo{title}{Magnetism}}, vol.~\bibinfo{volume}{4}
  (\bibinfo{publisher}{Academic Press, New York}, \bibinfo{year}{1966}).

\end{thebibliography}

\end{document}